\documentclass[a4paper,
               keeplastbox,   
               ]{jacow}
\usepackage{pdfpages,multirow,ragged2e} %
\newcommand{\lambdabarrr}{{\mkern0.75mu\mathchar '26\mkern -9.75mu\lambda}}
\makeatletter%
	\ifboolexpr{bool{xetex}}
	 {\renewcommand{\Gin@extensions}{.pdf,%
	                    .png,.jpg,.bmp,.pict,.tif,.psd,.mac,.sga,.tga,.gif,%
	                    .eps,.ps,%
	                    }}{}
\makeatother

\ifboolexpr{bool{xetex} or bool{luatex}} 
 {}                                      
 {\usepackage[utf8]{inputenc}}           

\usepackage[USenglish]{babel}

\ifboolexpr{bool{jacowbiblatex}}%
 {%
  \addbibresource{jacow-test.bib}
  \addbibresource{biblatex-examples.bib}
 }{}
\begin{document}

\title{Ultimate limits of future colliders\thanks{Work supported by DOE}}

\author{M.Bai \thanks{mbai@slac.stanford.edu}, SLAC National Accelerator Laboratory, Menlo Park, CA 94025, USA \\
V.Shiltsev\textsuperscript{1}, Fermilab, Batavia, IL 60510, USA \\
F. Zimmermann \textsuperscript{1}, CERN, 1211 Meyrin, Switzerland }
\maketitle

\begin{abstract}
   With seven operational colliders in the world and two under construction, the international particle physics community not only actively explores options for the next facilities for detailed studies of the Higgs/electroweak physics and beyond-the-LHC energy frontier, but also seeks a clear picture of the limits of the colliding beams method. In this paper, we try to consolidate various recent efforts in identifying physics limits of colliders in conjunction with societal sustainability, and share our thoughts about the perspective of reaching the ultimate quantum limit. 
\end{abstract}

\section{The Landscape of colliders}
\begin{figure}[!htb]
   \centering
   \includegraphics*[scale=0.22]{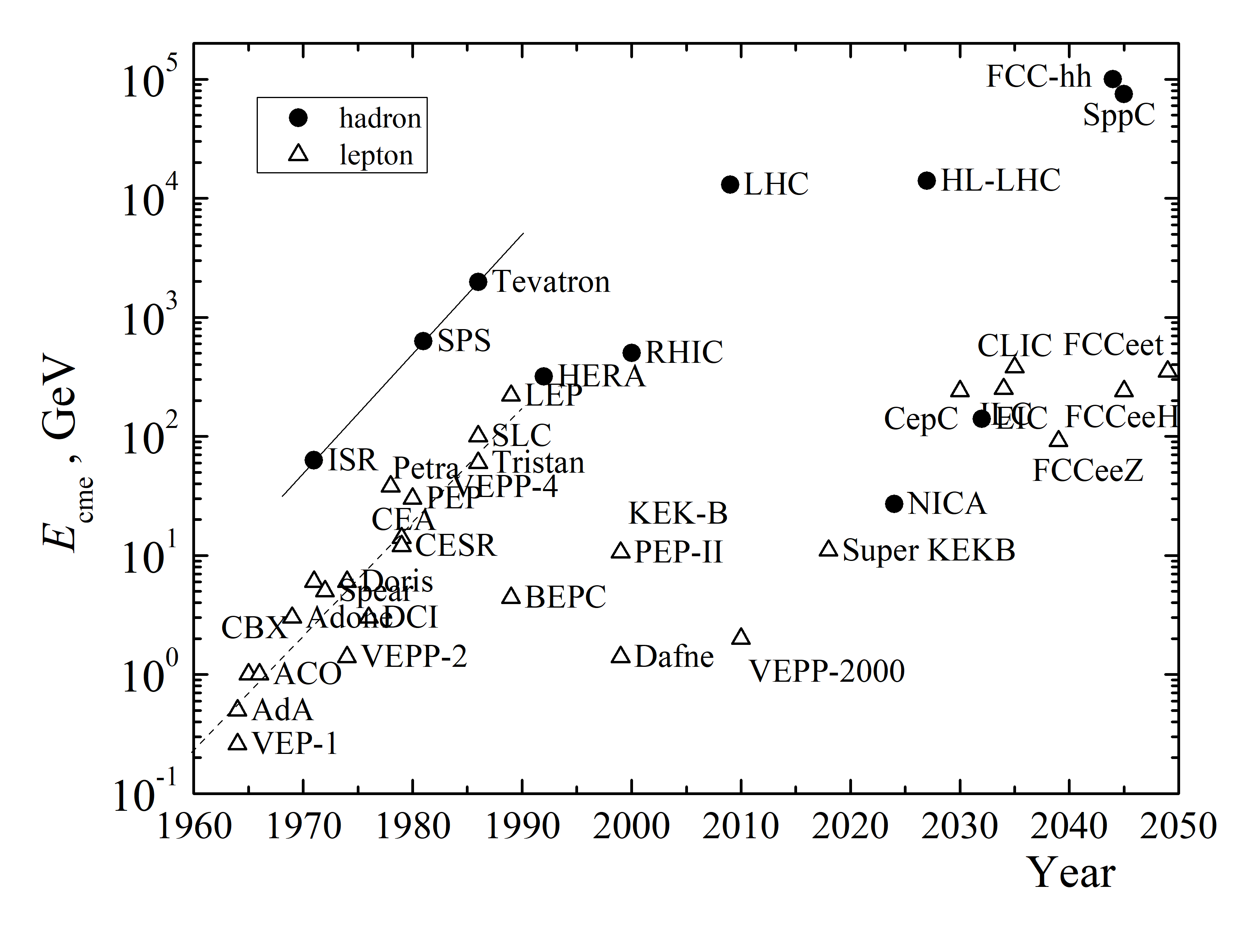}
   \caption{The so-called Livingston plot represents the evolution of the colliders from the past to the future \cite{rmp}.}
   \label{livingston}
\end{figure}

\begin{figure}[!htb]
   \centering
   \includegraphics*[scale=0.25]{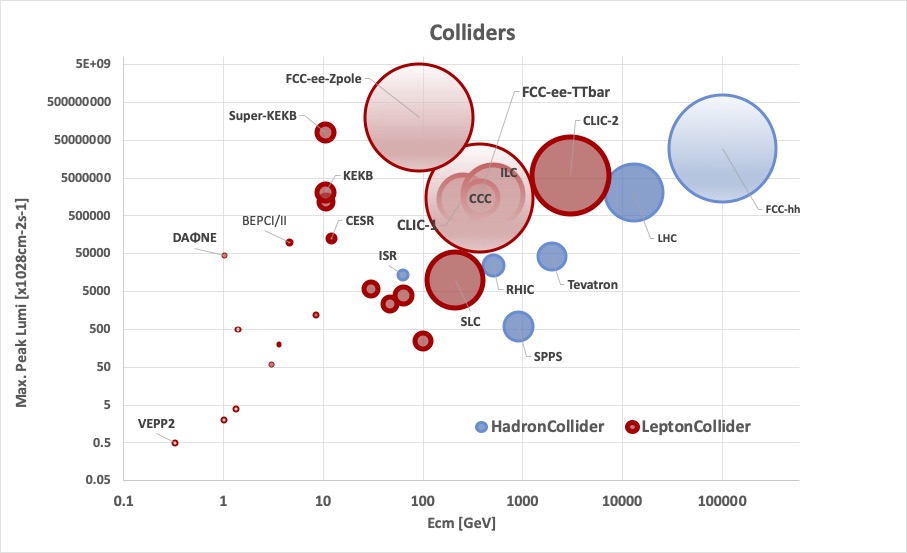}
   \caption{Collider peak luminosity and size as a function of centre-of-mass energy. The red bubbles represent the lepton colliders while the blue bubbles are the hadron colliders. For both cases, the semi transparent bubbles are the proposed colliders for future. The size (diameter or length) of the collider is represented by the size of the bubble.}
   \label{colliders}
\end{figure}

The development of accelerators and beams in the past century has led to incredible discoveries in physics, chemistry, biology, etc. Up to date, about 25 Nobel Prizes in Physics and 7 in Chemistry were made possible thanks to significant contributions from particles accelerators and beams~\cite{achao, rmp}. Among the family of accelerators, the 
collider has been the most important engine of discovery for particle physicists to produce new particles and to understand the fundamental laws that govern the subatomic structure. Figure~\ref{livingston} shows how the energy of colliders has increased orders of magnitudes over the past half a century.
Table~\ref{operatingcolliders} lists the colliders that are currently in operation and two which are under construction (NICA in Russia and the Electron Ion Collider at BNL in the USA).  
\begin{table}[!hbt]
   \centering
   \caption{Operational particle colliders.}
   \begin{tabular}{lccc}
       \toprule
       \textbf{Collider} & \textbf{Location} & \textbf{Species} & \textbf{ Energy per} \\
            & &  & \textbf{beam [GeV]}\\
       \midrule
           LHC/   & CERN,  &p     & 6800   \\ 
            \;    HL-LHC  & Europe &Pb, Xe &2560 \\
           RHIC  & BNL, USA      &p$^\uparrow$  &  255\\
                 &                &d, Au, etc & 100/n\\
     SuperKEKB      & KEK, Japan     &e$^+$e$^-$ & 7 e$^-$, 4 e$^+$     \\ 
        BEPCII      & IHEP, China    & e$^+$e$^-$       &  2.45      \\ 
        DA$\Phi$NE  & LNF, Italy          & e$^+$e$^-$      & 0.51        \\
        VEPP-2000  &  BINP, Russia      &    e$^+$e$^-$   & 1      \\
        VEPP-4M    & BINP, Russia     &  e$^+$e$^-$      & 6       \\
        NICA        & JINR, Russia  & Au, etc & 4.5/n        \\
        EIC         &BNL, USA  &e$^\uparrow$, p$^\uparrow$ & 18e$^-$, 255p\\
                    &          &He3   & 167/n\\
                    &          &Au. etc & 100/n \\
    \bottomrule
   \end{tabular}
   \label{operatingcolliders}
\end{table}
Despite a noticeable slow-down in the increase of the energy frontier  over the past couple of decades, the quest for further pushing the collider frontier has never abated. At the latest US High Energy Physics long term strategy community process known as {\it Snowmass'21}, the high energy physics community has composed its long-term energy-frontier road map, which consists of three elements: to ensure the success of the ongoing LHC luminosity upgrade (HL-LHC), to realize Higgs factory e$^+$e$^-$ collider for studying the Higgs boson and electroweak physics with a high precision, and to develop multi-TeV colliders for probing the 10 TeV parton energy scale  \cite{plub_liantao}. These requirements clearly push the future colliders, once again, into an hitherto unprecedented scale as shown in Fig.~\ref{colliders}, where they may face many challenges both in terms of reaching the advertised performance as well as for maintaining societal support~\cite{colliderlimits_VS}. 

\section{Ultimate limits of future energy frontier colliders}
\subsection{Energy}
For circular electron-positron colliders, due to the synchrotron radiation, the energy of the circular collider is ultimately limited by $E_{e^+e^-} \lesssim 500 {\rm GeV} (\frac{R}{10 {\rm km}})^{1/3}$, where $E_{e^+e^-}$ is the centre-of-mass energy and $R$ is the ring radius. 
The beam energy of a circular hadron collider is limited both by
the size $R$, and also by the available bending magnetic field.
The current proposed 100~TeV center-of-mass energy proton proton colliders
such as the FCC-hh at CERN, with a circumference close to 100 km, requires dipole magnets of 16~Tesla or beyond, which are currently under research and development. Larger colliders which have been suggested include the Collider in the Sea \cite{CS} and a Circular Collider on the Moon \cite{CCM}.  
However, the construction and operation of the future colliders of any type must also stay sustainable. 
To reach 1 PeV centre-of-mass energy in the FCC-hh's 91 km tunnel will require 10 times stronger bending magnet. To go further beyond, the circular proton proton collider will also experience a similar energy limit and size scaling as the electron-positron circular colliders, namely $E_{pp}<10 {\rm PeV} (\frac{R}{10 {\rm km}})^{1/3}$, due to synchrotron radiation.

For linear e$^+$s$^-$ colliders, the ultimate beam energy is determined by the limit of acceleration gradient. For CLIC with 100~MV/m acceleration gradient normal conducting technology, a 50 km long laser straight tunnel is required to reach 3~TeV. The recent development in cryogenic copper C-band technology offers a potential path to push the acceleration gradient towards 150 MV/m~\cite{sami}. Based on this technology, the newly proposed Cool Copper Collider~\cite{c3} concept could possibly enable a center-of-mass energy to 4.5~TeV within the same 50 km tunnel. To obtain centre-of-mass energies of 10~TeV and beyond will require quite a few substantial developments in
\begin{itemize}
    \item extreme high gradient RF acceleration structures; 
    \item laser straight long tunnel construction to avoid vertical emittance growth due to bending fields otherwise required to follow the earth curvature~\cite{ipac21}; and
    \item cost-effective high efficient RF power sources.
\end{itemize}

The concept of colliding muons, which emit much less synchrotron radiation then electrons, has been developed since several decades. Despite the advantage of being less susceptible to synchrotron radiation in comparison to electrons or positrons, the muon collider energy is ultimately limited by the available acceleration within the lifetime of the muons~\cite{colliderlimits_VS, ipac21}. Depending on the site of future multi-TeV muon collider, the neutrino flux from the muon decay is likely to become another factor limiting the maximum energy~\cite{colliderlimits_VS, ipac21}. 

\subsection{luminosity}
In general, the luminosity of a collider, assuming symmetric collision, is given by 
\begin{equation}
  L=
\frac{1}{2\pi}\frac{P_{wall}{\eta}}{E_{cm}}\frac{N_b}{\sigma_x^*\sigma_y^*} F_{geom} \; ,
  \label{lumi}
\end{equation}
where $P_{wall}$ denotes the wall-plug power, $\eta$ the ratio of wall-plug power to beam power, $N_b$ the number of particles per bunch, $\sigma_x^*$ and $\sigma_y^*$ the horizontal and vertical rms beam sizes at the interaction point (IP). Finally, $F_{geom}$ describes the geometric factors such as hour-glass, pinch effect for collisions of particles with opposite charge state, etc. 

Equation~(\ref{lumi}) shows that for a given optics at the IP, the luminosity scales linearly with energy. On the other hand, the discovery of new fundamental constituents with the lepton collider requires that the luminosity scales with the square of the centre-of-mass energy $E_{cm}$ as $\left( \frac{E_{cm}}{10TeV} \right)^2\times 10^{35}$~cm$^{-2}$s$^{-1}$~\cite{plub_liantao}. As an example, the minimum luminosity for a direct search of new physics at a lepton collider of 10 TeV $E_{cm}$ is $2\times10^{31}$~cm$^{-2}$s$^{-1}$ and minimum luminosity is $E_{cm}$ is $3\times10^{34}$~cm$^{-2}$s$^{-1}$ for precision measurement. 
However, extremely high luminosity at lower energies can also lead to discoveries through the observation of rare processes.
For example, operating the FCC-ee as a ``TeraZ'' factory at 91 GeV c.m.~could lead to the discovery of sterile right-handed neutrinos.

For hadron colliders, the mass of the particle that can be created scales with $E_{cm}^{2/3}L^{1/6}$~\cite{lee,rmp, plub_liantao}, and the ultimate aim, in this case, is to reach collision at $\ge$ 100 TeV centre-of-mass energy with $10^{35} - 10^{36}$~cm$^{-2}$s$^{-1}$ luminosity~\cite{plub_liantao}. 
Hence, the beam brightness needs to be scaled with the energy accordingly for future colliders. This requires one to overcome many limiting factors starting from obtaining and preserving low emittance bright beams through the acceleration process, final focusing as well as the impact of beam-beam interaction to beam dynamics and detector background. Thanks to the advancements in beam dynamics and beam techniques, such as 
compensation of various beam-beam effects, crab crossing, etc, over the past decades, the luminosity performance of current operating colliders has been impressive, as is shown in Fig.~\ref{colliders}. Despite their low energy and compact size, both DA$\Phi$NE and BEPC-II were able to reach much higher peak luminosity than earlier generations of circular colliders at the same energies. Also LEP, CESR and the three B factories PEP-II, KEKB, SuperKEKB achieved many orders of magnitude higher luminosity than the first generation high energy e$^+$e$^-$ linear collider SLC. Nevertheless, to reach similar peak luminosity performance, future multi-TeV e$^+$e$^-$ colliders face additional unique limiting factors due to beamstrahlung, Oide effect, and coherent pair production. These effects have been carefully studied and are well understood. Also various mitigation measures have been proposed. 

In principle, the ultimate achievable normalized beam emittance in each of three dimensions for a bunch of $N_b$ unpolarized fermions (electrons or positrons) is its quantum limit, i.e.~$\gamma\epsilon_n^{QM}={N_b}^{1/3}\lambdabarrr_p/4$, where $\lambdabarrr_p$ is the particle’s de Broglie
wavelength~\cite{qm}. 
In such a scenario, without taking into account  other limiting factors such as beamstrahlung and Oide effect, the ultimately quantum-limited luminosity is still orders of magnitude above the luminosity targets desired for the multi-TeV e$^+$e$^-$  colliders~\cite{ipac21}. 
Nevertheless, utilization of a beam with a quantum-limited emittance at the collision point at multi TeV energies 
requires a beam delivery system able to manipulate beams of extremely small dimensions, much below the current state of the art. 
\subsection{Advanced acceleration concept based collider}
\begin{figure}[!htb]
   \centering
   \includegraphics*[scale=0.12]{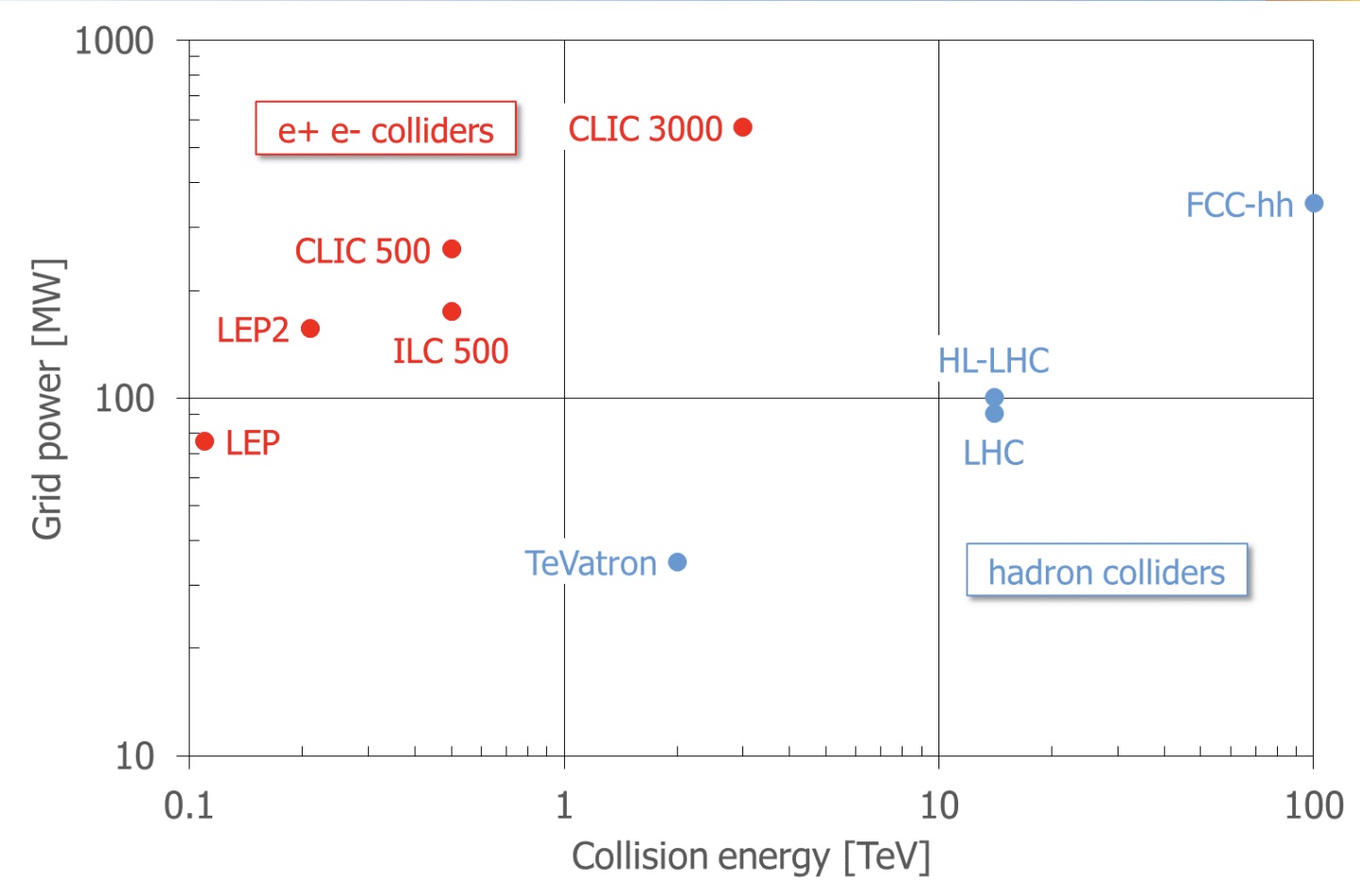}
   \caption{Power consumption of current and proposed future energy frontier colliders~\cite{lebrun}.}
   \label{power}
\end{figure}

It is evident that, so far, the colliders based on conventional RF technology have not yet reached their physical limits, neither in energy nor in luminosity. Nevertheless, the actual physical engineering limits and societal sustainability have been the two main limiting factors of the proposed future high energy colliders. Figure~\ref{power} shows the power consumption of energy frontier colliders, both currently in operation and proposed for the future~\cite{lebrun}. For the case of LHC, the current power consumption of LHC operation including detectors, but without the rest of the CERN accelerator complex, is 0.6 TWh per year, which is about 10\% of what Swissgrid produces~\cite{lhc_fact}. Assuming the same running scenario, FCC could be consuming close to 50\% what Swissgrid produces.
Hence, disruptive technologies for acceleration as well as for beam manipulation, such as high temperature superconducting technologies, may be required to find a path forward to realize the future energy frontier colliders within the limits of societal sustainability, and to continue pushing the energy frontier far beyond the currently proposed multi-TeV colliders. 

Both laser driven and beam driven plasma wakefield acceleration, a.k.a.~LWFA and PWFA,  have been pursued and intensified worldwide. While unprecedented acceleration gradient has been demonstrated with both PWFA and LWFA, the achieved beam quality, both intensity and brightness, has not yet reached a comparable level to what conventional RF technology based colliders have routinely delivered. The recent progress of AWAKE has shown encouraging steps towards a possibly very high energy electron-proton collider based on beam-driven PWFA~\cite{plub_awake}. Nevertheless, the projected luminosity is on the order of $10^{28}$ to $10^{29}$ $cm^{-2}s^{-1}$, and falls far short of the physics requirement. 
Overall, the path towards next generation TeV colliders still requires numerous new marvels in beam physics as well as in engineering to meet the repetition rate, staging requirement and ultimately the beam performance that conventional RF technology based accelerators have always achieved.

Nevertheless, as the advanced concept acceleration field is rapidly developing, it is not appropriate to estimate the performance limit at this point of time. In addition to accelerating gradient, beam quality and energy efficiency will be of high importance.

In addition to the LWFA and PWFA, other creative ideas such as using crystals for accelerating and finally focusing and colliding positively charged particles, or accelerating single particle with black hole to reach Planck energy have been proposed~\cite{plub_chen, brooks}. For a crystal channeling collider, the estimated luminosity is~\cite{plub_chen}
\begin{equation}
    L_c=f_{rep}n_{ch}\frac{N_c^2\sqrt{KE}}{2\pi\hbar c}
\end{equation}
where, $n_{ch}$ is number of crystal channels and $N_c$ is the number of particles captured in each crystal channel, $K$ is the crystal's channeling strength.  Typical crystal $K$ is around $10^{11}$~GeV/m$^2$. As the focusing in the crystal accelerator is continuous, quantum emittance could be achieved with particles remaining in the ground state. While these ideas are still at very infant stage, the crystal channeling technique (without longitudinal acceleration) has been successfully developed and used for collimation at LHC.

Other potentially disruptive concepts include the collision of crystalline beams \cite{cb1,cb2}, 
and harnessing quantum entanglement for acceleration or focusing \cite{brooks}.

\section{CONCLUSION}
The past half a century has seen the rise of colliders that led to many discoveries. While the pace for new colliders at the energy frontier has slowed down, the quest for the next discovery collider still remains strong. The realization of currently proposed future colliders, circular or linear e$^+$e$^-$ colliders and 100 TeV pp colliders, is likely to push the energy frontier by another order of magnitude. Nevertheless, both in terms of energy and luminosity, these machines are still far below the ultimate frontier set by the quantum limit. 

Hence, new paradigms for acceleration and beam manipulation are required to overcome the barriers that seem to be formidable with currently available technologies. 
A breakthrough in this domain will not only benefit the energy frontier particle physics, but it can also be a game changer for other accelerator based scientific fields such as those served by X-ray Free Electron Lasers.

\section{ACKNOWLEDGEMENTS}
The authors would like to thank the co-organizers and participants of the {\it Physics Limit of Ultimate Beams} workshop series.
%
%
\ifboolexpr{bool{jacowbiblatex}}%
	{\printbibliography}%
	{%
	

} 
%
%


\end{document}